\documentclass[superscriptaddress,twocolumn,10pt, prl,aps,showpacs]{revtex4-2}

\usepackage{fancyhdr} 
\usepackage[caption=false, position=top,textfont=normalfont,singlelinecheck=off,justification=raggedright]{subfig}
\usepackage{tabularx} 
\usepackage{graphicx} 
\usepackage{float}
\usepackage[table]{xcolor} 
\usepackage{dsfont} 
\usepackage{physics} 
\newcommand{\brab}[1]{\langle{#1}\rVert}
\newcommand{\ketb}[1]{\lVert{#1}\rangle}
\newcommand{\braketb}[2]{\langle  {#1} \lVert {#2}  \rangle}

\usepackage[normalem]{ulem}
\usepackage{url}
\usepackage[pdfpagelabels=false,
 breaklinks=true,
 hypertexnames=false,
 colorlinks=true,
 linkcolor={blue},
 citecolor={red},
 urlcolor={blue},
 anchorcolor={black}]{hyperref} 

\usepackage{orcidlink}

\makeatletter
\def\maketitle{
\@author@finish
\title@column\titleblock@produce
\suppressfloats[t]}
\makeatother

\begin{document}

\author{David S. Schlegel\,\orcidlink{0000-0003-2013-1676}}
\email{david.schlegel@epfl.ch}
\affiliation{Laboratory of Theoretical Physics of Nanosystems (LTPN), Institute of Physics, École Polytechnique Fédérale de Lausanne (EPFL), CH-1015 Lausanne, Switzerland}
 \affiliation{Center for Quantum Science and Engineering, École Polytechnique Fédérale de Lausanne (EPFL), CH-1015 Lausanne, Switzerland}
\author{Fabrizio Minganti\,\orcidlink{0000-0003-4850-1130}}
\affiliation{Laboratory of Theoretical Physics of Nanosystems (LTPN), Institute of Physics, École Polytechnique Fédérale de Lausanne (EPFL), CH-1015 Lausanne, Switzerland}
\affiliation{Center for Quantum Science and Engineering, École Polytechnique Fédérale de Lausanne (EPFL), CH-1015 Lausanne, Switzerland}
\author{Vincenzo Savona\,\orcidlink{0000-0002-8984-6584}}
\affiliation{Laboratory of Theoretical Physics of Nanosystems (LTPN), Institute of Physics, École Polytechnique Fédérale de Lausanne (EPFL), CH-1015 Lausanne, Switzerland}
\affiliation{Center for Quantum Science and Engineering, École Polytechnique Fédérale de Lausanne (EPFL), CH-1015 Lausanne, Switzerland}

\title{Coherent-State Ladder Time-Dependent Variational Principle \texorpdfstring{\\for Open Quantum Systems}{}}

\begin{abstract}
We present a new paradigm for the dynamical simulation of interacting many-boson open quantum systems. The method relies on a variational ansatz for the $n$-boson density matrix, in terms of a superposition of photon-added coherent states. It is most efficient for the simulation of driven-dissipative systems where the state is well described by quantum fluctuations on top of a displaced field, making it suitable for the simulation of several coupled modes with large occupation numbers, that are otherwise very challenging using a Fock-space expansion. We test our method on several examples, demonstrating its potential application to the predictive simulation of interacting bosonic systems and cat qubits.
\end{abstract}

\date{\today}

\maketitle
\paragraph{Introduction.---}
Quantum technologies rely on the generation and control of many-body quantum states on platforms that maintain long-lived quantum coherence~\cite{SchoelkopfNature2008, BlochNature2012, MonroeScience2013}.
While quantum simulators and computers are expected to eventually outperform classical emulation~\cite{CiracNature2012, BernienNature2017, TacchinoAdvancedQuantumTechnologies2020, CiracNanoPhotonics2021}, the certification and verification of quantum protocols still require efficient approximate methods to simulate entangled many-body dynamics \cite{MontanaroNature2016, PreskillNISQ2018}. 
This task is especially challenging for the many bosonic degrees of freedom characterizing superconducting circuits~\cite{AruteNat19, Kjaergaard20}, integrated photonic platforms \cite{wang_integrated_2020,pelucchi_potential_2022}, optomechanical resonators~\cite{AspelmeyerRMP2014}, and trapped ions~\cite{BruzewiczApplPhysRev2019} among others.

In quantum optics, several methods have been developed to model weakly-interacting systems
close to their classical limit, based on the Gross-Pitaevskii equation (GPE)~\cite{gross_structure_1961, Pitaevskii19961, Carusotto_RMP_2013_quantum_fluids_light} or the Gutzwiller mean-field approximation \cite{BoitePRL2013, LeePRL2013, JinPRX2016, CasteelsPRA18, HuybrechtsPRA20}.
Interactions, however, can induce quantum correlations that cannot be captured by a mean-field approximation, nor by methods such as the truncated Wigner approximation or low-order cumulant expansion~\cite{Feist2020,VicentiniPRA18}.
Similarly, driven bosonic systems can operate in a regime where the state inherits a large degree of coherence from the engineered driving fields. A prototypical example are bosonic quantum codes, where drive and dissipation create and stabilize quantum superpositions of coherent states~\cite{OfekNat16, RosenblumScience2018, SivakNature2023,JoshiQST21}.
The prohibitively large Hilbert space of multimode, driven bosonic systems is often an obstacle to the use of conventional simulation techniques based on Fock-space truncation~\cite{Fock1932, ReedSimon1975}. 
Novel methods, such as the \emph{shifted Fock basis}~\cite{ChamberlandNoh2022}, mitigate this problem when simulating the steady state of the system subject to a stationary driving field.
The efficient numerical simulation of the dynamics of driven-dissipative many-body bosonic systems remains, however, a major challenge.

In this Letter, we model the dynamics of a many-body driven-dissipative bosonic system using the time-dependent variational principle~\cite{DiracPrinciples1982, FrenkelWaveMechanics1934, McLachLanMolecularPhysics1964} (TDVP). We introduce an ansatz based on a set of unnormalized coherent states called a \textit{coherent-state ladder}. These states are chosen so to have optimal overlap with the corner of the Hilbert space spanned by the system dynamics. 

The method naturally accounts for symmetries, such as rotational symmetries in phase space characterizing, e.g., bosonic cat codes~\cite{MirrahimiNJP14, OfekNat16, AlbertPRA2018, GuillaudPRX2019, PuriScienceAdvances2020, ChamberlandNoh2022, GautierArXiV2023} and other rotation symmetric bosonic codes~\cite{SchlegelPRA2022, GrimsmoPRX2020}, thus permitting an efficient study of time-dependent gate protocols and the occurrence of errors.
The number of basis states enters as a controllable parameter that sets the accuracy of the approximation. It allows simulating coupled interacting bosonic systems of considerable size requiring only moderate computational resources.

Hereunder, we detail the principle of the method and benchmark its performance by applying it to various models of driven-dissipative bosonic systems. Our method outperforms the Fock-space truncation method in all examples, requiring fewer basis states to represent the system's dynamics accurately.

\begin{figure*}[t]
\subfloat[]{\includegraphics[width=\columnwidth]{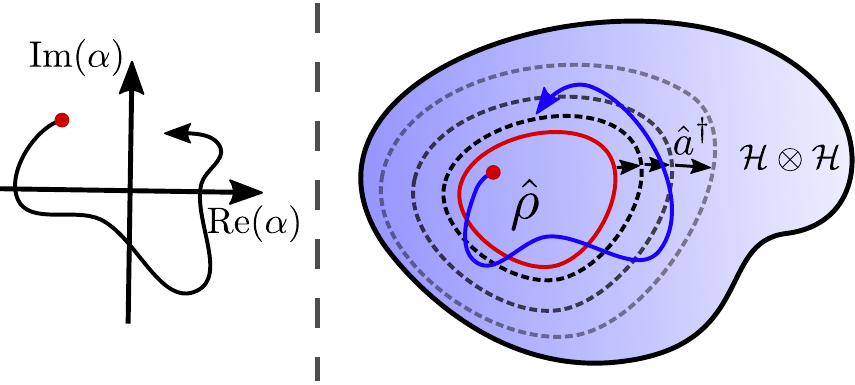}}
\hfill
  \subfloat[]{
    \begingroup
\renewcommand{\arraystretch}{1.6} 
    \begin{tabularx}{\columnwidth}[b]{l|X|X|X|X}
      \textbf{method}  & \textbf{large $\hat{n}$} & \textbf{inter-actions} &\textbf{symme-tries} & \textbf{dyna-mics} \\ \hline
      Fock-space~\cite{Fock1932, ReedSimon1975} & \cellcolor{red!25}no & \cellcolor{green!25}yes & \cellcolor{green!25}yes & \cellcolor{green!25}yes\\ \hline
      GPE~\cite{gross_structure_1961, Pitaevskii19961} & \cellcolor{green!25}yes & \cellcolor{red!25}no & \cellcolor{red!25}no & \cellcolor{green!25}yes \\ \hline
      shifted Fock~\cite{ChamberlandNoh2022} & \cellcolor{green!25}yes & \cellcolor{green!25}yes & \cellcolor{green!25}yes & \cellcolor{red!25}no \\ \hline
      our method & \cellcolor{green!25}yes & \cellcolor{green!25}yes & \cellcolor{green!25}yes & \cellcolor{green!25}yes
    \end{tabularx}
    \endgroup
    \label{subtbl:the-table}
  }

\caption{(a) Schematic illustration of the variational ansatz. 
The TDVP equations of motion in Eqs.~(\ref{eq:rhodot_multimode}-\ref{eq:alphadot_multimodeb}) couple the coherent-field amplitude $\alpha$ (left) of the Bargmann state $\ketb{\alpha}$ [c.f. Eq.~\eqref{eq:BargmannDefinition}] and the density matrix elements $B_{ij}$ of the co-moving basis $\mathcal{S}_N = \{\ketb{\alpha}, \hat{a}^\dagger \ketb{\alpha}, {{}\hat{a}^\dagger}^2 \ketb{\alpha} \dots {{}\hat{a}^\dagger}^N \ketb{\alpha}\}$. (b) Comparison of various methods with respect to the ability to capture large occupation number $n$, interactions, rotational symmetries, and the dynamics of an open bosonic quantum system.}
\label{fig:Fig1}
\end{figure*}

\paragraph{Method.---}
We consider a bosonic open quantum system coupled to a Markovian bath \cite{BreuerBookOpen}, whose dynamics is governed by the Lindblad master equation
\begin{equation}
    \label{eq:Lindblad_general}
    \dv{\hat{\rho}}{t} = -i \left[\hat{H}, \hat{\rho}\right] + \sum\limits_j \gamma_j \mathcal{D}\left[\hat{L}_j\right](\hat{\rho}).
\end{equation}
Here, $\hat{H}$ is the Hamiltonian and  $\mathcal{D}[\hat{L}_j](\hat{\rho})$ are the dissipators, with jump-operators $\hat{L}_j$ and dissipation rates $\gamma_j$, defined by \mbox{$\mathcal{D}[\hat{A}](\hat{\rho}) \equiv \hat{A}\hat{\rho}\hat{A}^\dagger - \frac{1}{2} \hat{A}^\dagger\hat{A}\hat{\rho} - \frac{1}{2}\hat{\rho}\hat{A}^\dagger \hat{A}$}.

To implement the TDVP, we adopt the formalism proposed in Ref.~\cite{Joubert-DoriolJournalChemicalPhysics2015}. The ansatz for the density matrix is then
\begin{equation}\label{eq:rhoansatz}
\hat{\rho} = \sum_{i,j}^N B_{ij}(t) \ketbra{\phi_i(t)}{\phi_j(t)},
\end{equation}
where the states $\ket{\phi_{i}(t)}$ are not in general orthogonal, and are expressed in terms of variational parameters. The matrix elements $B_{ij}$ also enters as additional variational parameters. Once a criterion for determining the states $\ket{\phi_{i}(t)}$ is introduced, the number $N$ will determine the expressive power of the ansatz and the accuracy of the variational approach.

The parametrized states $\ket{\phi_{i}(t)}$ that we introduce are based on the \emph{Bargmann} states $\ketb{\alpha}$~\cite{BargmannAppliedMathematics1961, Segal1967, Perez2000}, i.e., non-normalized coherent states defined by
\begin{equation}\label{eq:BargmannDefinition}
    \ketb{\alpha} = e^{\alpha \hat{a}^\dagger} \ket{0} = \sum\limits_{n=0}^\infty \frac{\alpha^n}{\sqrt{n!}} \ket{n} = e^\frac{|\alpha|^2}{2} \ket{\alpha},
\end{equation}
with $\ket{n}$ being Fock states.
Starting from $\ketb{\alpha}$, the subsequent application of the ladder operator $\hat{a}^\dagger$ yields the (non-orthonormal) photon-added coherent-state ladder  basis $\mathcal{S}_N = \{\ketb{\alpha}, \hat{a}^\dagger \ketb{\alpha}, {{}\hat{a}^\dagger}^2 \ketb{\alpha} \dots {{}\hat{a}^\dagger}^N \ketb{\alpha}\}$~\cite{AgarwalPRA1991}.
For $\ketb{\alpha = 0}$, $\mathcal{S}_N $ is just the Fock (number) basis.
For $\alpha\ne0$, $\mathcal{S}_N $ describes excitations of the state $\ketb{\alpha}$. 

The states $\mathcal{S}_N$ bring several advantages in the 
simulation of the dynamics.
First, they express efficiently the state of a driven system, which is expected to be given by a coherent state plus small quantum fluctuations.
Furthermore, the Bargmann states obey:
\begin{equation}
\hat{a}\ketb{\alpha} = \alpha \ketb{\alpha},\quad 
\pdv{\ketb{\alpha}}{\alpha} = \hat{a}^\dagger \ketb{\alpha}.\label{eq:BargmannDerivative}    
\end{equation}
From Eq.~\eqref{eq:BargmannDerivative} it follows that the tangent space of $\mathcal{S}_N$ lies in $\mathcal{S}_{N+1}$. In TDVP, the tangent space contains the gradient function, which dictates the updates of the variational parameters, whose evaluation is therefore greatly simplified. Notice also that Bargmann states are holomorphic functions of the coherent-field amplitude $\alpha$, which further simplifies the TDVP formulation using McLachlan's principle~\cite{BroeckhoveChemPhysLett1988, KanPRA1981}.

As depicted schematically in Fig.~\ref{fig:Fig1}, we propose to use the McLachlan's variational principle~\cite{McLachLanMolecularPhysics1964, BroeckhoveChemPhysLett1988, Raab2000} on a multi-mode bosonic system, by extending the density matrix ansatz in Eq.~\eqref{eq:rhoansatz} to $M$ bosonic modes. Each mode is expressed by its coherent-field amplitude $\alpha_k$~\cite{Joubert-DoriolJournalChemicalPhysics2015} and by the related coherent-state ladder $\mathcal{S}_{N}^{(k)} = \{\hat{a}_k^{n}\ketb{\alpha}, n = 0 \dots N\}$, where $k=1,\,\ldots,\,M$. 
The TDVP translates then into coupled differential equations for the variational parameters $\boldsymbol{B} = \{ B_{i\, j}\}$ and $\alpha_k$ (See Supplemental Material for details)
\begin{align}\label{eq:rhodot_multimode}
        \boldsymbol{\dot{B}} &= \boldsymbol{S}^{-1} \boldsymbol{L}\boldsymbol{S}^{-1} -\boldsymbol{S}^{-1}\boldsymbol{\tau}\boldsymbol{B} - \boldsymbol{B} \boldsymbol{\tau}^\dagger \boldsymbol{S}^{-1},\\
        \dot{\alpha_k} &=  \frac{\text{Tr}\{\boldsymbol{Y_0}^{(k)} \boldsymbol{B}\}}{\text{Tr}\{\boldsymbol{C_0}^{(k)}[\boldsymbol{B}\boldsymbol{S}\boldsymbol{B}]\}},\label{eq:alphadot_multimodeb}
\end{align} 
where the matrices $\boldsymbol{S}$, $\boldsymbol{L}$, $\boldsymbol{C}$, $\boldsymbol{\tau}$, $\boldsymbol{Y_0}^{(k)}$ and $\boldsymbol{C_0}^{(k)}$ depend on $\alpha_k$, $\dot{\alpha}_k$, and the model parameters.

\paragraph{Application Ia: Driven-dissipative Kerr resonator.---} As a first proof of principle, we consider a single system consisting of a single-photon drive and Kerr-nonlinearity, whose Hamiltonian reads
\begin{equation}
    \hat{H} = \frac{U}{2}{{}\hat{a}^\dagger}^2\hat{a}^2 +  F(\hat{a} + \hat{a}^\dagger),
\end{equation}
with single-photon driving amplitude $F$ and Kerr-nonlinearity $U$.
The system is subject to single-photon loss $\kappa\mathcal{D}[\hat{a}]$.
In the classical limit of $U\rightarrow 0$ and $F\rightarrow\infty$, with $F\sqrt{U}$ constant, the required number of Fock-states needed to accurately describe the systems' dynamics increases when decreasing $U$, as the average photon number of the system diverges as $(F/U)^{2/3}$~\cite{BartoloPRA16}.
In Fig.~\ref{fig:Fig2}(a), we show the number of required basis states needed to reach a fidelity $\mathcal{F}>99\%$, both for the Fock basis and our TDVP basis. Our results show that while the number of Fock states increases by lowering $U$, the required number of basis states for our TDVP method remains nearly constant.
In Fig.~\ref{fig:Fig2}(b), we show the infidelity $1-\mathcal{F}$ as a function of time for different basis sizes $N$, starting from the steady-state predicted by the semiclassical approximation. The results illustrate how the quality of the TDVP solution increases with $N$.

\paragraph{Application Ib: Asymmetrically driven nonlinear photonic dimer.---}
\begin{figure*}
    \centering
    \includegraphics[width=\textwidth]{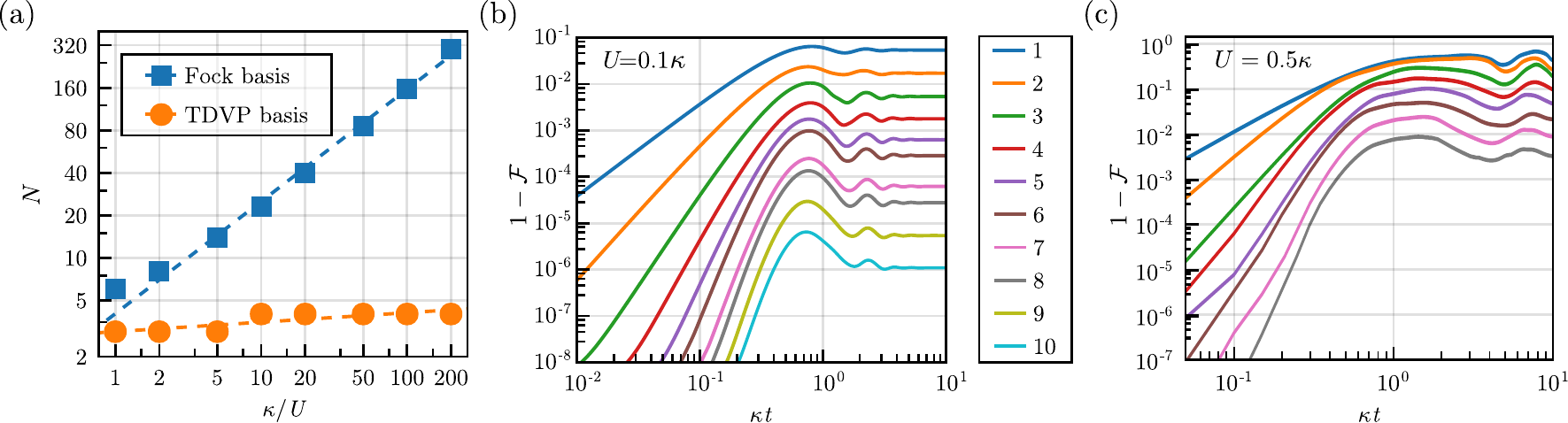}
    \caption{
    (a) Number of basis states $N$ needed to keep the fidelity above the value $\mathcal{F}=0.99$ along the dynamics of the dissipative Kerr resonator for the Fock-basis (blue) and the coherent-ladder method (orange) for $F = 1.5\sqrt{\kappa^3/U}$. (b-c) Infidelity $1-\mathcal{F}$ as a function of dimensionless time $\kappa t$, for (b) different numbers of basis states $N$ in the dissipative Kerr resonator and (c) in the asymmetrically driven nonlinear photonic dimer for $J=1.2\kappa$, $\Delta_{1,2}=2\kappa$, $\alpha_1(t=0) =  - 1.0 - 1.84i$, $\alpha_2(t=0)=1.36 + 0.8i$.}
    \label{fig:Fig2}
\end{figure*}

Second, we consider a system of two coupled modes with Hamiltonian
\begin{equation}
\begin{aligned}
\hat{H}= & \sum_{i=1,2}-\Delta \hat{a}_i^{\dagger} \hat{a}_i+\frac{U}{2} {{}\hat{a}_i^{\dagger}}^2 \hat{a}_i^2 \\
& -J\left(\hat{a}_1^{\dagger} \hat{a}_2+\hat{a}_1 \hat{a}_2^{\dagger}\right)+F\left(\hat{a}_1^{\dagger}+\hat{a}_1\right),
\end{aligned}
\end{equation} 
and single-photon loss processes $\kappa \mathcal{D}[\hat{a}_1]$ and $\kappa \mathcal{D}[\hat{a}_2]$. Here, $\Delta$ is the detuning of the resonators with respect to the drive frequency, $J$ the hopping interaction, and $F$ the driving-field amplitude, acting only on one of the two modes.
A semiclassical treatment using the GPE predicts multiple parametrically unstable regions for different parameter regimes~\cite{Sarchi_coherent_2008,SeiboldPRA20}.
Again, in the classical limit of $U\to0$ and $F\to\infty$, with $F\sqrt{U}$ constant, accurately describing the effect of quantum fluctuations with a truncated-Fock approach becomes challenging as $|\alpha|^2$, and thus the average number of bosons per mode, diverge as $(F/U)^{2/3}$~\cite{BartoloPRA16}.
Furthermore, the field amplitude $\alpha$ varies periodically along the dynamics, and cannot be modeled using a stationary ansatz, like e.g. with the shifted Fock basis~\cite{ChamberlandNoh2022}.
This system, therefore, provides an ideal testbed for our method.

Starting from an initial coherent state, we simulate the time-evolution for different basis sizes $N$.
The results are shown in Fig.~\ref{fig:Fig2}(c), where the infidelity $1-\mathcal{F}$, with respect to the exact solution, is plotted.
Our results indicate that with few basis states in the variational ansatz, the essential dynamics of the system can be accurately captured.

\paragraph{Rotational symmetries.---}
Our approach conveniently facilitates the inclusion of rotational symmetries, thereby enabling the modeling of rotation-symmetric quantum codes such as cat codes.
As an example, we consider a $\mathds{Z}_2$ rotational symmetry whose symmetry subspaces can be decomposed in even and odd superpositions of Bargmann states with opposite displacement, yielding the unnormalized even and odd cat states
\begin{equation}\label{eq:Bargmanncat}
\ketb{\mathcal{C}_{\alpha}^\pm} = \ketb{\alpha} \pm \ketb{-\alpha}.
\end{equation}
Similar to the previous case, we can construct a cat-ladder basis comprising the two symmetry sectors by applying $\hat{a}^\dagger$ on the even and odd cat states in Eq.~\eqref{eq:Bargmanncat}.
Note that $\pdv{\ketb{\mathcal{C}_{\alpha}^\pm}}{\alpha} = \hat{a}^\dagger \ketb{\mathcal{C}_{\alpha}^\mp}$.
Equations~\eqref{eq:rhodot_multimode}~and~\eqref{eq:alphadot_multimodeb} maintain the same structure, with slight modifications in the involved matrices~(See Supplemental Material for details).
A $\mathds{Z}_N$ rotational symmetry can similarly be encoded by defining a basis of superpositions of rotated coherent Bargman states $e^{i k 2\pi \hat{n}/N}\ketb{\alpha}$, with \mbox{$k=0,\,\dots,\,N-1$}. In the following, we apply our method to systems of interacting cat qubits.

\paragraph{Application IIa: Dynamics of a quenched cat qubit.---} Let us consider the dynamics of a cat qubit, stabilized by the simultaneous action of two-photon dissipation and Kerr nonlinearity when the drive is quenched. The system is described by the Hamiltonian
\begin{equation}\label{eq:Kerr_resonator}
\hat{H} = \left(\frac{G(t)}{2} \hat{a}^2 + \frac{G(t)^*}{2} \hat{a}^{\dagger 2}\right) + \frac{U}{2} {{}\hat{a}^{\dagger}}^2 \hat{a}^2,
\end{equation}
and dissipator $\eta\mathcal{D}[\hat{a}^2]$, where $G(t)=G_0$ for $t<0$ and $G(t)=G_1$ for $t>0$ is the two-photon drive amplitude and $\eta$ the two-photon dissipation rate.
We suppose that the system reached its steady state at $t<0$, and we investigate the dynamics after the quench at $t=0$.
\begin{figure*}
    \includegraphics[width=\textwidth]{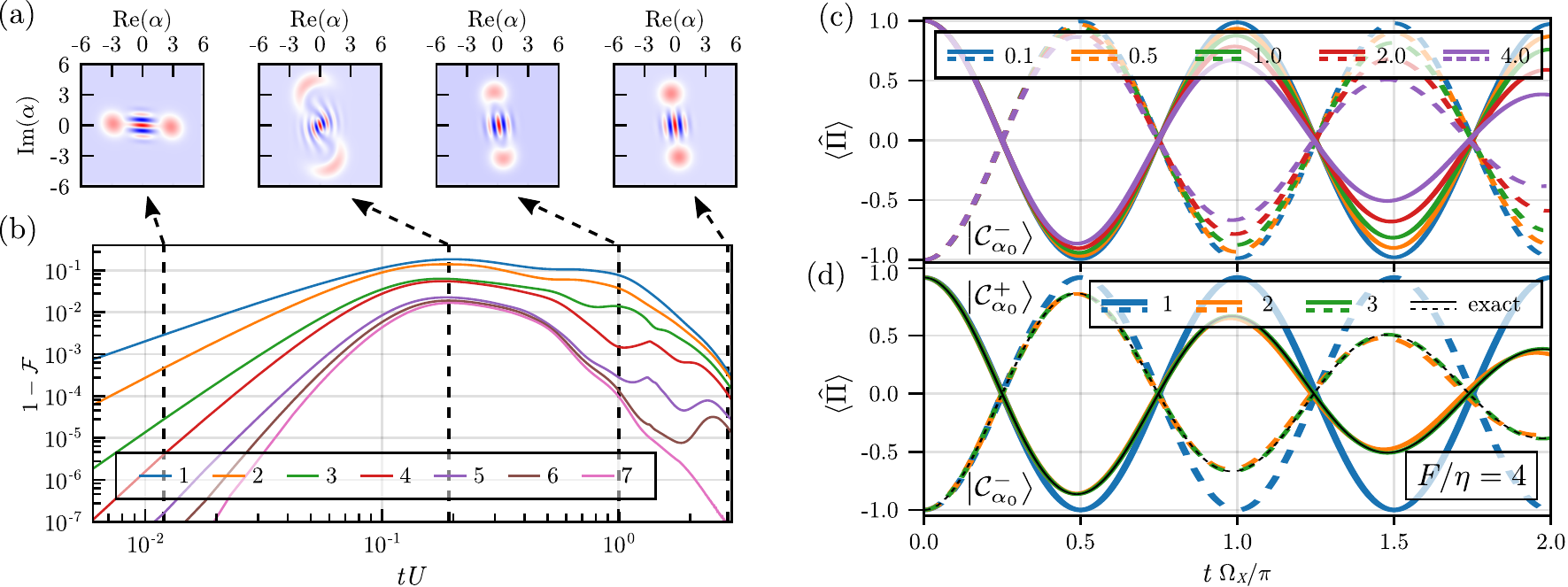}
    \caption{(a) Snapshots of the Wigner function at different times along the dynamics for an initial cat state, highlighting leakage out of the cat-state manifold during the transient dynamics for $\eta = U/4$, $G_0=5iU$, $G_1=-i5U$. (b) Infidelity $1-\mathcal{F}$ as a function of time for different basis sizes $N$. (c) Parity expectation value $\expval*{\hat{\Pi}}$ as a function of $t \Omega_X /\pi$ for different ratios $F/\eta$, assuming an initial even cat state (solid lines) and odd cat state (dashed lines), with $G=10\eta$. (d) For a fixed ratio $F/\eta = 4$, the quantity $\expval*{\hat{\Pi}}$ is plotted for different basis sizes $N$.}
    \label{fig:Fig3}
\end{figure*}
In \mbox{Fig.~\ref{fig:Fig3}(a)}, we depict snapshots of the Wigner functions along the time evolution.
The leakage in the transient dynamics appears as a departure from the shape of an ideal cat state.
We show the effect of the basis cutoff $N$ in Fig.~\ref{fig:Fig3}(b).
Despite the large leakage out of the cat-qubit manifold, a relatively small basis size captures the whole dynamics with high accuracy.

\paragraph{Application IIb: Coherent non-adiabatic rotation of a cat qubit.---}
We consider now the case $U=0$, where the steady-state is spanned by the cat-qubit manifold with amplitude $\alpha = \sqrt{iG/{\eta}}$. We encode $\ket{{0/1}_\mathrm{L}}\equiv \ket*{\mathcal{C}_\alpha^\pm}$ as the logical $\hat{Z}$-eigenstates. These states are eigenstates of the photon number parity $\hat{\Pi} = (-1)^{\hat{n}}$.
A coherent rotation around the logical $\hat{X}$-axis can be realized by a single-photon drive \mbox{$\hat{H}_{X} = F(\hat{a} + \hat{a}^\dagger)$}~\cite{MirrahimiNJP14}, with single-photon driving amplitude $F$. In the limit $F \ll \eta $, the code space remains unperturbed, and the qubit is adiabatically rotated by an angle $\varphi = \Omega_{X}t$, with Rabi frequency $\Omega_{X} = \sqrt{2} F|\alpha|$~\cite{MirrahimiNJP14}.
To analyze the non-adiabatic effect of the rotation-gate $\hat{H}_{X}$, we simulate the dynamics of the system starting from $\ket*{\mathcal{C}_\alpha^\pm}$.
In Fig.~\ref{fig:Fig3}(c), we show the average parity $\expval*{\hat{\Pi}}$ for different single-photon driving amplitudes $F$ as a function of time.
For $F/\eta \gtrsim 1$, the single-photon drive induces non-adiabatic processes, resulting in code leakage and a reduction of the fidelity of the rotation gate.
To assess the efficiency of the method, we plot in \mbox{Fig.~\ref{fig:Fig3}(d)} the parity $\expval*{\hat{\Pi}}$ for different basis sizes for a fixed driving strength $F$.
Again, our results reveal that even for a moderate truncation, the non-adiabatic nature of the $\hat{X}$-rotation gate can be adequately described, significantly reducing the computational overhead compared to a truncated Fock approach.

\paragraph{Application III: Interacting cat qubit dynamics.---}
We now consider a two-mode cat-like bosonic system.
Each mode ($\hat{a}_1$ and $\hat{a}_2$) is described by a two-photon driven-dissipative Kerr resonator analogous to Eq.~\eqref{eq:Kerr_resonator} that interacts through the hopping (beam-splitter) Hamiltonian
\begin{equation}\label{eq:interactionHamiltonian}
    \hat{H}_\mathrm{int} = \sum\limits_{\langle i, j \rangle}J_{ij}(\hat{a}_i \hat{a}_j^\dagger + \hat{a}_i^\dagger \hat{a}_j),
\end{equation} with hopping interaction $J_{ij}$.
The system conserves the total photon number parity $\expval*{\hat{\Pi}_1\hat{\Pi}_2}$~\cite{AlbertPRA14}. 
The coherent tunneling interaction $\hat{H}_\mathrm{int}$ enables coherent exchange of excitations between the two modes, thereby encoding an entangling two-qubit gate.
We simulate the time evolution of the system starting from an initial product state $\ket{\psi(0)} = \ket*{\mathcal{C}_{\alpha_1}^+} \otimes \ket*{\mathcal{C}_{\alpha_2}^+}$, where $\alpha_1$ and $\alpha_2$ differ from those characterizing the steady state of the cat code.
This may represent a system where the drive changes upon application of the $\hat{X}_1\hat{X}_2$ gate.
We compute the time evolution of the system,  assuming different basis sizes $N$, and compare to the exact result in \mbox{Fig.~\ref{fig:Fig4}(a)}, where we show the evolution of $\expval*{\hat{a}^2}$ and the single-mode parity $\expval*{\hat{\Pi}_1}$ in \mbox{Fig.~\ref{fig:Fig4}(b)}. Convergence to the exact result arises already for small basis size $N$ per mode.
\begin{figure*}[t]
 \includegraphics[width=\textwidth]{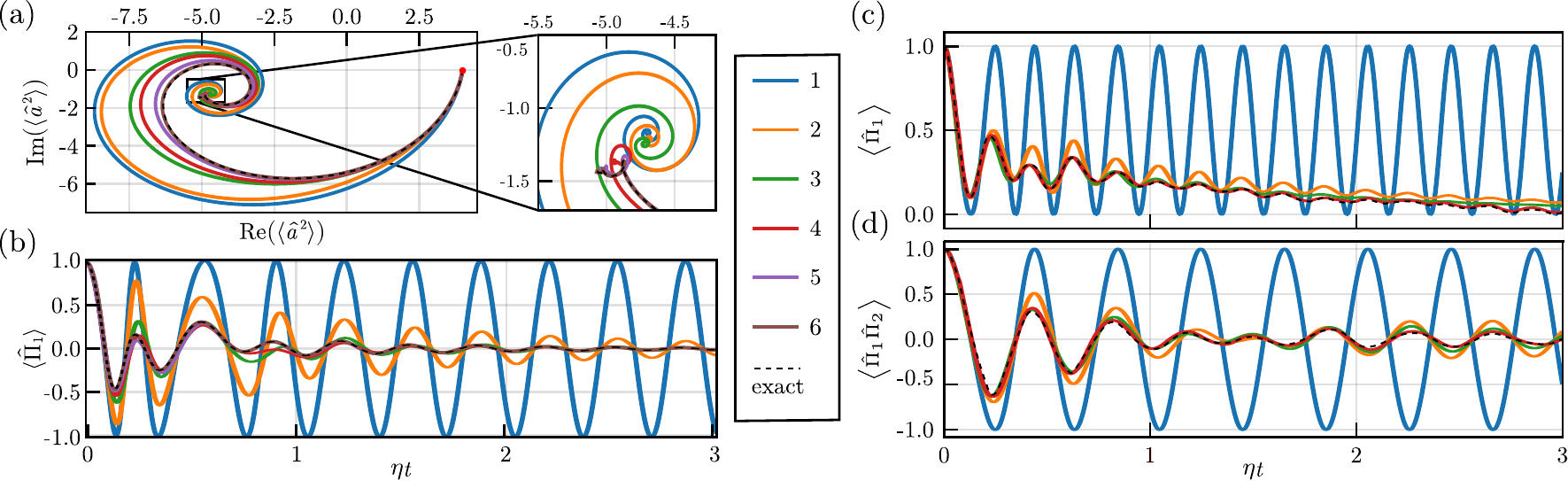}
 \caption{(a) Real and imaginary parts of $\expval*{\hat{a}^2}$, as computed along the dynamics of a system of two interacting driven-dissipative Kerr-resonators. (b) Single-mode parity $\expval*{\hat{\Pi}_1}$ as a function of time for the same system, setting $G_i= 5U,\; \eta_i = 0.25U, \; J_{12} = U$ and initial state:~\mbox{$\ket{\psi(0)}  = \ket*{\mathcal{C}_{\alpha_1}^+,\mathcal{C}_{\alpha_2}^+} $}, with $\alpha_1 = \alpha_2 = 2$. (c)-(d) Single-mode parity $\expval*{\hat{\Pi}_1}$ and two-mode parity $\expval*{\hat{\Pi}_1\hat{\Pi}_2}$ as a function of time, as computed for a system of three coupled modes, setting $G_i= 5U\; \eta_i = 0.25U, \; J_{12} = J_{23} = 0.8U$ and initial state $\ket{\psi(0)}  = \ket*{\mathcal{C}_{\alpha_1}^+, \mathcal{C}_{\alpha_2}^+,\mathcal{C}_{\alpha_3}^+} $, with $\alpha_i = 2$.}
 \label{fig:Fig4}
\end{figure*}
Finally, we simulate a system consisting of three interacting bosonic modes arranged in an open chain configuration, where the nearest neighbor hopping constants $J_{12}$ and $J_{23}$ are non-zero. In the regime we consider, simulating this system with a truncated Fock approach starts being computationally challenging. Convergence vs $N$ is then assessed by simply comparing results for increasing $N$.
Figs.~\ref{fig:Fig4}(c-d) display the time evolution of the single-mode parity $\expval*{\hat{\Pi}_1}$ and two-mode parity $\expval*{\hat{\Pi}_1 \hat{\Pi}_2}$ for various single-mode basis sizes $N$. Analogously to the two-mode case, the total photon number parity $\expval*{\hat{\Pi}_1 \hat{\Pi}_2 \hat{\Pi}_3}$ is preserved throughout the system dynamics. Even for the three-mode case, convergence is reached within a reasonable number of basis states, highlighting the potential of the method.

\paragraph{Conclusions.---}
We have introduced an approach for simulating the dynamics of interacting many-boson driven-dissipative systems. The coherent-state ladder approach employs a variational ansatz for the $n$-boson density matrix, where both the matrix elements and the displacement fields serve as time-dependent variational parameters. This approach bridges the gap between quantum and classical realms, making it suitable for simulating systems with large occupation numbers that are beyond the capabilities of Fock-space methods.

We have demonstrated the potential of the method through several examples, showcasing its applicability to interacting bosonic systems and cat qubits. It outperforms Fock-space methods in terms of the number of basis states needed to accurately represent the system's dynamics. 

Our method can straightforwardly be extended to include different parametrizations of the basis, such as squeezed states or higher-order processes, as well as other types of symmetries (See Supplemental Material for details).
It can thus accurately study the system dynamics of multi-mode bosonic systems in the presence of quantum correlations.

The coherent-state ladder thus offers new possibilities for accurately studying system dynamics of a wide range of single- and multi-mode bosonic systems, whose simulation would be otherwise very challenging. It opens new pathways in the simulation of systems of many coupled bosonic qubits to efficiently optimize gate and error-correction protocols~\cite{GuillaudPRX2019, PuriScienceAdvances2020, ChamberlandNoh2022, GautierArXiV2023}.

\begin{acknowledgments}
\paragraph{Acknowledgments.---}
This work was supported by the Swiss National Science Foundation through Projects No. 200020\_185015 and 200020\_215172 and was conducted with the financial support of the EPFL Science Seed Fund 2021.
\end{acknowledgments}

\bibliography{biblio}

\cleardoublepage\phantomsection
\setcounter{page}{1}

\title{Supplemental Material:\texorpdfstring{\\Coherent-State Ladder Time-Dependent Variational Principle}{}
\texorpdfstring{\\for Open Quantum Systems}{}}
\maketitle

\pagenumbering{roman}
\pagestyle{fancy}

\fancyhf{}

\rhead{Supplemental Material \thepage}

\section{Derivation of the TDVP equations of motion}
\label{appendix:TDVPEOMderivations}
In the McLachlan TDVP for density matrices, the equations of motion arise from the minimization of the Frobenius norm of the error~\cite{McLachLanMolecularPhysics1964, BroeckhoveChemPhysLett1988}
\begin{equation}
    \left\| \dot{\hat{\rho}} - \mathcal{L}[\hat{\rho}] \right\| = \text{Tr}\{(\dot{\hat{\rho}} - \mathcal{L}[\hat{\rho}])^\dagger(\dot{\hat{\rho}} - \mathcal{L}[\hat{\rho}])\}^{\frac{1}{2}},
\end{equation}
which is equivalent to the stationary condition~\cite{Raab2000}
\begin{equation}\label{eq:MacLachlantdvp}
    \text{Tr}\{\delta\hat{\rho}(\dot{\hat{\rho}} - \mathcal{L}[\hat{\rho}])\} = 0,
\end{equation}
with an infinitesimal variation in the density matrix $\delta\hat{\rho}$.
For our particular choice of ansatz in Eq.~\eqref{eq:rhoansatz}, infinitesimal changes in the density matrix $\hat{\rho}$ translate to the variational paramters as
\begin{equation}\label{eq:rho_derivative}
    \delta \hat{\rho} = \sum\limits_{ij}\pdv{\hat{\rho}}{B_{ij}} \delta B_{ij} 
    + \sum\limits_k\pdv{\hat{\rho}}{\alpha_k} \delta \alpha_k + \sum\limits_k\pdv{\hat{\rho}}{\alpha_k^*} \delta \alpha_k^*.
\end{equation}
Inserting the above expression in Eq.~\eqref{eq:MacLachlantdvp}, and due to the independent nature of the derivatives, we have three stationary conditions:
\begin{align}
    \text{Tr}\left\{{\pdv{\hat{\rho}}{B_{ij}}} (\dot{\hat{\rho}} - \mathcal{L}[\hat{\rho}])\right\} &=  0 \label{eq:tdvp_rho}\\
    \text{Tr}\left\{{\pdv{\hat{\rho}}{\alpha_k}} (\dot{\hat{\rho}} - \mathcal{L}[\hat{\rho}])\right\} &=0\label{eq:tdvp_alpha}\\
    \text{Tr}\left\{{\pdv{\hat{\rho}}{\alpha_k^*}} (\dot{\hat{\rho}} - \mathcal{L}[\hat{\rho}])\right\} &=0 \label{eq:tdvp_alphastar}
\end{align}
Since the last two are complex conjugates of one another, they constitute the same condition.
We define the overlap matrix
\begin{equation}
    [\boldsymbol{S}]_{ij} = \braket{\phi_i}{\phi_j}.
\end{equation}
As the coherent-state- and cat-ladder basis states are not orthogonal, $\boldsymbol{S}$ is not an orthogonal matrix.
Note that due to this non-orthonormal basis, the projector $\hat{P}$, projecting onto the ansatz basis $\mathcal{S}$, is given by
\begin{equation}\label{eq:projector}
    \hat{P} = \sum\limits_{ij} \ket{\phi_i} [\boldsymbol{S}^{-1}]_{ij} \bra{\phi_j}.
\end{equation}

We can rewrite Eq.~\eqref{eq:rho_derivative} as
\begin{equation}\label{eq:rhodot_ansatz}
    \dv{\hat{\rho}}{t} = \sum\limits_{cd} \dot{B}_{cd} \ketbra*{\phi_c}{\phi_d} + 
    B_{cd} \ketbra*{\dot{\phi_c}}{\phi_d} + B_{cd}\ketbra*{\phi_c}{\dot{\phi_d}}
\end{equation}

Replacing this expression into Eq.~\eqref{eq:tdvp_rho} yields:
\begin{align}
    &\text{Tr}\left\{\pdv{\hat{\rho}}{B_{ij}}\left(\dot{\hat{\rho}} - \mathcal{L}[\hat{\rho}]\right)\right\} = 0\\
    &= \sum\limits_{kl} \bra{\phi_l} \boldsymbol{S}_{kl}^{-1} \left[
    \ketbra{\phi_i}{\phi_j} \left(\sum\limits_{cd}\dot{B}_{cd} \ketbra*{\phi_c}{\phi_d}  \right. \right.\\
    &\qquad \left. \left.+ 
    B_{cd} \ketbra*{\dot{\phi_c}}{\phi_d} + B_{cd}\ketbra*{\phi_c}{\dot{\phi_d}} - \mathcal{L}[\hat{\rho}] \right)
    \right] \ket{\phi_k} \nonumber \\
    &=\sum\limits_{cd} S_{jc}\dot{B}_{cd}S_{di} + \tau_{jc}B_{cd} S_{di}  + S_{jc}B_{cd}\tau_{id}^* - L_{ji},
\end{align}
where we have introduced
\begin{align}
    [\boldsymbol{\tau}]_{mn} &= \sum\limits_{k}\braket{\phi_m}{\pdv{\phi_n}{\alpha_k}} \dot{\alpha_k},\\
    [\boldsymbol{L}]_{ij} &= \mel{\phi_i}{\mathcal{L}[\hat{\rho}]}{\phi_j}.
\end{align}

We thus have the condition
\begin{equation}
 \boldsymbol{S}\boldsymbol{\dot{B}}\boldsymbol{S} + \boldsymbol{\tau} \boldsymbol{B} \boldsymbol{S} + \boldsymbol{S}\boldsymbol{B}\boldsymbol{\tau}^\dagger - \boldsymbol{L} = 0,\label{eq:tdvp_rho_matrixform_implicit}
\end{equation} which gives the equation of motion for the density matrix coefficients,
\begin{equation}\label{eq:eom_rho}
    \boldsymbol{\dot{B}} = \boldsymbol{S}^{-1} \boldsymbol{L}\boldsymbol{S}^{-1} -\boldsymbol{S}^{-1}\boldsymbol{\tau}\boldsymbol{B} - \boldsymbol{B} \boldsymbol{\tau}^\dagger \boldsymbol{S}^{-1}.
\end{equation}

For the equation of motion for $\dot{\alpha}_k$, we have by inserting the definition of $\dot{\hat{\rho}}$ from Eq.~\eqref{eq:rhodot_ansatz},
\begin{align}
    & \text{Tr}\left\{{\pdv{\hat{\rho}}{\alpha_k^*}} (\dot{\hat{\rho}} - \mathcal{L}[\hat{\rho}])\right\} = 0\label{eq:app_alpha_1}\\
    \begin{split}\label{eq:app_alpha_2}
    &= \sum\limits_{ab}\bra{\partial_{\alpha_k} \phi_b}
    \left\{ \sum\limits_{cd} \left( \dot{B}_{cd} \ketbra*{\phi_c}{\phi_d} \right. \right. \\
    &\qquad + \left. \left.
    B_{cd} \ketbra*{\dot{\phi_c}}{\phi_d} + B_{cd}\ketbra*{\phi_c}{\dot{\phi_d}}
    \right) - \mathcal{{L}}[\hat{\rho}]\right\}\ket{\phi_a}B_{ab}.
    \end{split}
\end{align}
    Inserting the definition of $\dot{\boldsymbol{B}}$ from Eq.~\eqref{eq:eom_rho}, we obtain
\begin{align}
    &\text{Tr}\left\{{\pdv{\hat{\rho}}{\alpha_k^*}} (\dot{\hat{\rho}} - \mathcal{L}[\hat{\rho}])\right\}\\
    \begin{split}\label{eq:app_alpha_3}
    &= \sum\limits_{ab}\bra{\partial_{\alpha_k} \phi_b}
    \Big\{ \sum\limits_{cd} \big( [\boldsymbol{S}^{-1} \boldsymbol{L}\boldsymbol{S}^{-1}   \\ 
    &\qquad \qquad \qquad -\boldsymbol{S}^{-1}\boldsymbol{\tau}\boldsymbol{B} - \boldsymbol{B} \boldsymbol{\tau}^\dagger \boldsymbol{S}^{-1}]_{cd} \ketbra*{\phi_c}{\phi_d}\\
    &\qquad + 
    B_{cd} \ketbra*{\dot{\phi_c}}{\phi_d} + B_{cd}\ketbra*{\phi_c}{\dot{\phi_d}}
    \big) - \mathcal{{L}}[\hat{\rho}]\Big\}\ket{\phi_a}B_{ab}
    \end{split}
\end{align}
By using the definition of the projector $\hat{P}$ in Eq.~\eqref{eq:projector}, we have
\begin{align}
    &\text{Tr}\left\{{\pdv{\hat{\rho}}{\alpha_k^*}} (\dot{\hat{\rho}} - \mathcal{L}[\hat{\rho}])\right\}\\
    \begin{split}\label{eq:app_alpha_4}
    &= \sum\limits_{ab}\bra{\partial_{\alpha_k} \phi_b}
    \Big\{ \sum\limits_{cd} \big(\hat{\mathds{1}} - \hat{P} \big) \ket*{\dot{\phi}_c}B_{cd}S_{da}\\
    &\qquad \qquad
    +    \big(\hat{P} - \hat{\mathds{1}} \big) \mathcal{{L}}[\hat{\rho}] \ket{\phi_a}\Big\} B_{ab}
    \end{split}\\
    \begin{split}\label{eq:app_alpha_5}
    &=  \sum\limits_{abcl}\bra{\partial_{\alpha_k} \phi_b}
    [\hat{\mathds{1}} - \hat{P} ] \ket*{\partial_{\alpha_l}\phi_c} [\boldsymbol{B}\boldsymbol{S}\boldsymbol{B}]_{cb} \dot{\alpha}_l\\
    &\qquad \qquad
    -   \bra{\partial_{\alpha_k} \phi_b} [\hat{\mathds{1}} - \hat{P}] \mathcal{{L}}[\hat{\rho}] \ket{\phi_a}B_{ab}
    \end{split},
    \end{align}
where in the last step we have inserted the definition of $\ket*{\dot{\phi}_c}$.
Importantly, since $\bra{\partial_{\alpha_k} \phi_m}
[\hat{\mathds{1}} - \hat{P} ] \ket*{\partial_{\alpha_l}\phi_n} = 0$ for $k\neq l$, as a variation of $\alpha_k$ only affects states in mode $k$, we can drop the sum over $l$ in Eq.~\eqref{eq:app_alpha_5}.
Defining
\begin{align}
    [\boldsymbol{C_0}]_{mn}^{k} &= \bra{\partial_{\alpha_k} \phi_m}
    [\hat{\mathds{1}} - \hat{P} ] \ket*{\partial_{\alpha_k}\phi_n},\label{eq:app:C0}\\
    [\boldsymbol{Y_0}]_{mn}^k &= \bra{\partial_{\alpha_k} \phi_m} [\hat{\mathds{1}} - \hat{P}] \mathcal{{L}}[\hat{\rho}] \ket{\phi_n},\label{eq:app:Y0}
\end{align}
we thus obtain
    \begin{align}
    &\text{Tr}\left\{{\pdv{\hat{\rho}}{\alpha_k^*}} (\dot{\hat{\rho}} - \mathcal{L}[\hat{\rho}])\right\}\\
    &= \sum\limits_{abc} \boldsymbol{C_0}_{bc}^{k} [\boldsymbol{B}\boldsymbol{S}\boldsymbol{B}]_{cb} \dot{\alpha}_k
    - \boldsymbol{Y_0}_{ba}^k \boldsymbol{B}_{ab}\label{eq:app_alpha_6}\\
    &= \text{Tr}\{\boldsymbol{C_0}^{k} [\boldsymbol{B}\boldsymbol{S}\boldsymbol{B}]\} \dot{\alpha}_k
    - \text{Tr}\{\boldsymbol{Y_0}^k \boldsymbol{B}\}.\label{eq:app_alpha_7}
\end{align}
    
Finally, the equation of motion for $\dot{\alpha}_k$ reads:
\begin{equation}
    \dot{\alpha}_k = \frac{\text{Tr}\{\boldsymbol{Y_0}^k \boldsymbol{B}\}}{\text{Tr}\{\boldsymbol{C_0}^{k} [\boldsymbol{B}\boldsymbol{S}\boldsymbol{B}]\}}.
\end{equation}
We hence have an implicit ordinary differential equation defined by Eqs.~\eqref{eq:rhodot_multimode}, \eqref{eq:alphadot_multimodeb}, reading
\begin{align}
        \boldsymbol{\dot{B}} &= \boldsymbol{S}^{-1} \boldsymbol{L}\boldsymbol{S}^{-1} -\boldsymbol{S}^{-1}\boldsymbol{\tau}\boldsymbol{B} - \boldsymbol{B} \boldsymbol{\tau}^\dagger \boldsymbol{S}^{-1},\\
        \dot{\alpha_k} &=  \frac{\text{Tr}\{\boldsymbol{Y_0}^{(k)} \boldsymbol{B}\}}{\text{Tr}\{\boldsymbol{C_0}^{(k)}[\boldsymbol{B}\boldsymbol{S}\boldsymbol{B}]\}}.
\end{align}

\section{Operator decomposition}
\label{appendix:Operatordecomposition}
We now detail the decomposition of operators and the construction of the overlap matrix for the coherent-state ladder basis and the cat-ladder basis.
\subsection{Coherent-state ladder}
For the coherent-state ladder with $\mathcal{S} = \{\ketb{\alpha,n} = {{}\hat{a}^\dagger}^n\ketb{\alpha}, n = 0 \dots N \}$, we can represent combinations of creation and annihilation operators as:
\begin{equation}
\begin{split}
    \hat{a} \ketb{\alpha, n} &= \hat{a}{{}\hat{a}^\dagger}^n \ketb{\alpha} = ({{}\hat{a}^\dagger}^n \hat{a} + n {{}\hat{a}^\dagger}^{n-1}) \ketb{\alpha} \\ &=  \alpha \ketb{\alpha, n} + n \ketb{\alpha, n-1}.
    \end{split}
\end{equation}
Hence with the overlap matrix $\boldsymbol{S}$ with elements \mbox{$S_{mn} = \langle \alpha, m\rVert\alpha, n\rangle$}, we have
\begin{equation}
    \brab{\alpha, m}\hat{a}\ketb{\alpha, n} = \alpha S_{m,n} + n S_{m, n-1}.
\end{equation}
Representations of other operators are obtained similarly.

We have the following recursion relation for the overlap matrix $\boldsymbol{S}$:
\begin{align}
S_{mn} = & \brab{\alpha, m-1}\hat{a}\hat{a}^\dagger\ketb{\alpha, n-1}\label{eq:app:recursion-coherenta}\\
\begin{split}
 =& S_{m-1, n-1}+(m-1)(n-1) S_{m-2, n-2}\\
&+|\alpha|^2 S_{m-1, n-1}+\alpha(m-1) S_{m-2, n-1}\\
&+\alpha^*(n-1) S_{m-1, n-2}.
\end{split}\label{eq:app:recursion-coherentb}
\end{align}
This allows to recursively calculate the matrix elements $S_{mn}$.

\subsection{Cat ladder}
In the non-orthonormal basis of Bargmann cat states, we can decompose the annihilation operator $\hat{a}$ in the subsystem decomposition by calculating its action on a basis state:
\begin{equation}\label{eq:annihilation_op_decomp}
    \hat{a}\ketb{\mathcal{C}_{\alpha, n}^\pm} = \alpha \ketb{\mathcal{C}_{\alpha, n}^\mp} + n \ketb{\mathcal{C}_{\alpha, n-1}^\mp},
\end{equation} 
and hence in a modular subsystem decomposition~\cite{PanteloniPRL2020} the annihilation operator reads
\begin{equation}
   \hat{a} = \hat{X} \otimes (\tilde{b} + \alpha).
\end{equation}
With the inner product matrix
\begin{equation}
    S_{mn}^{\mu\nu} = \braketb{\mathcal{C}_{\alpha, m}^\mu}{\mathcal{C}_{\alpha, n}^\nu},
\end{equation}
where indices $\mu,\;\nu \in \{+,-\}$ label the parity of the state, we have
\begin{equation}
    \brab{\mathcal{C}_{\alpha, m}^\mu}\hat{a}\ketb{\mathcal{C}_{\alpha, n}^\nu} = \alpha S_{mn}^{\mu,\bar{\nu}} + n S_{m,n-1}^{\mu,\bar{\nu}},
\end{equation}where $\bar{\nu}$ indicates a flip of parity index $\nu$.
Analogous, we can obtain expressions for the representation of other operators in this basis.

In the cat-ladder basis, the recursion relation to calculate the inner product matrix elements $S_{mn}^{\mu\nu}$ is analogous to Eqs.~\eqref{eq:app:recursion-coherenta},\eqref{eq:app:recursion-coherentb}:
\begin{align}
S_{mn}^{\mu \nu } = & \brab{\mathcal{C}_{\alpha, m-1}^{\bar{\mu}}}\hat{a}\hat{a}^\dagger\ketb{\mathcal{C}_{\alpha, n-1}^{\bar{\nu}}}\\
\begin{split}
 =& S_{m-1, n-1}^{\bar{\mu}  \bar{\nu }}+(m-1)(n-1) S_{m-2, n-2}^{\mu  \nu }\\
&+|\alpha|^2 S_{m-1, n-1}^{\mu  \nu }+\alpha(m-1) S_{m-2, n-1}^{\mu  \nu }\\
&+\alpha^*(n-1) S_{m-1, n-2}^{\mu  \nu }.
\end{split}
\end{align}
With the overlap matrix for the cat-qubit
\begin{equation}
    S_{00}^{\mu\nu} = \begin{pmatrix}
    4 \cosh{\left(|\alpha|^2\right)} & 0 \\
    0 & 4 \sinh{\left(|\alpha|^2\right)}
    \end{pmatrix},
\end{equation}
$S_{mn}^{\mu \nu }$ can be recursively computed for $m, n>0$.

\section{Numerical implementation}
\label{appendix:numericalimplementation}
We here sketch a numerical implementation of the method presented in this paper.
We first consider a single bosonic mode in the cat-ladder basis.
Since calculating a first-order derivative with respect to $\alpha$ directly corresponds to the application of $\hat{a}^\dagger$ and parity conjugation in the chosen basis, we can easily relate derivatives to block shifts in the inner product matrix:
\begin{equation}
    \braketb{\mathcal{C}_{\alpha, m}^{\mu}}{\partial_\alpha \mathcal{C}_{\alpha, n}^{\nu}} = \brab{\mathcal{C}_{\alpha, m}^{\mu}}\hat{a}^\dagger\ketb{\mathcal{C}_{\alpha, n}^{\bar{\nu}}} = 
    S_{m, n+1}^{\mu, \nu}.
\end{equation}
We can then calculate the matrices $\mathbf{C}_0$ and $\mathbf{Y}_0$, defined by Eqs.~\eqref{eq:app:C0},~\eqref{eq:app:Y0} in terms of block-shifted matrices.
By denoting block-shifts of matrices in the form
$[\boldsymbol{A}^{(l)}]_{mn}^{\mu \nu} = [\boldsymbol{A}]_{m+1,n}^{\mu \nu}$, $[\boldsymbol{A}^{(r)}]_{mn}^{\mu \nu} = [\boldsymbol{A}]_{m,n+1}^{\mu \nu}$, and $[\boldsymbol{A}^{(l,r)}]_{mn}^{\mu \nu} = [\boldsymbol{A}]_{m+1,n+1}^{\mu \nu}$ for any Matrix $\boldsymbol{A}$ with elements $[\boldsymbol{A}]_{mn}^{\mu\nu}$, we can rewrite $\boldsymbol{C_0}$ and $\boldsymbol{Y_0}$ as
\begin{align}
\mathbf{C}_0 & =\boldsymbol{S}^{(l,r)}-\boldsymbol{S}^{(l)}\boldsymbol{S}^{-1}\boldsymbol{S}^{(r)} \label{eq:app:C0numeric}\\
\mathbf{Y}_0 & =\sum_p\left(\mathbf{A}^{(l)}_p \mathbf{B D}_p -\mathbf{S}^{(l)} \mathbf{S}^{-1} \mathbf{A}_p \mathbf{B D}_p\right) \\
& =\mathbf{L}^{(l)} -\mathbf{S}^{(l)} \mathbf{S}^{-1} \mathbf{L}.\label{eq:app:Y0numeric}
\end{align}
Here we have expanded the Liouvillian matrix $\boldsymbol{L}$ in its operator-sum representation, $\boldsymbol{L} = \sum_p \boldsymbol{A}_p \boldsymbol{B} \boldsymbol{D}_p$, with left and right Kraus operator matrices $\boldsymbol{A}_p$ and $\boldsymbol{D}_p$.

Note that matrix $\mathbf{C}_0$, which can be interpreted as a quantum-geometric tensor, is not full rank, but only has two diagonal non-zero elements associated with the highest moment of $\hat{a}^\dagger$ included in the basis (one for each parity sector).
Therefore, for any initial density matrix that has no population for the highest moment included in the basis $\text{Tr}\{\boldsymbol{C_0}[\boldsymbol{B}\boldsymbol{S}\boldsymbol{B}]\}$ will be zero.
As a result $\dot{\alpha}$ will be zero until the density matrix $\boldsymbol{B}$ populates the highest moment included in the basis ansatz.
Intuitively, this can be understood in the following way: For our time-dependent variational ansatz, the equations of motion describe a dynamics in which the density matrix $\boldsymbol{B}$ evolves into a full-rank matrix. Upon reaching the corner (i.e. the highest moment of $\hat{a}^\dagger$ included in the basis), the basis re-centers according to the evolution of $\alpha$.
In the numerical simulations, we therefore evolve $\alpha$ only if $|\text{Tr}\{\boldsymbol{C_0}[\boldsymbol{B}\boldsymbol{S}\boldsymbol{B}]\}| > \epsilon$, with a small threshold $\epsilon$, otherwise we set $\dot{\alpha}=0$ and evolve only $\boldsymbol{B}$.

For multiple bosonic modes, the numerical implementation can be straightforwardly extended. We first note that due to the choice of the basis, the inner product matrix $\boldsymbol{S}$ can be expressed as a tensor product of inner product matrices of each bosonic mode, i.e. $\boldsymbol{S} = \otimes_k \boldsymbol{S}_k$, also reducing the computation of the inverse of $\boldsymbol{S}$ to the inverses of $\boldsymbol{S}_k$.
Using block-shifts of matrices for mode $k$, we can express left and right derivatives of matrices with respect to $\alpha_k$.
Analogous to Eqs.~\eqref{eq:app:C0numeric},~\eqref{eq:app:Y0numeric}, we obtain
\begin{align}
\mathbf{C}_0^{(k)} & =\boldsymbol{S}^{(l_k,r_k)}-\boldsymbol{S}^{(l_k)}\boldsymbol{S}^{-1}\boldsymbol{S}^{(r_k)} \label{eq:app:C0knumeric}\\
\mathbf{Y}_0^{(k)} & =\sum_p\left(\mathbf{A}^{(l_k)}_p \mathbf{B D}_p -\mathbf{S}^{(l_k)} \mathbf{S}^{-1} \mathbf{A}_p \mathbf{B D}_p\right) \\
& =\mathbf{L}^{(l_k)} -\mathbf{S}^{(l_k)} \mathbf{S}^{-1} \mathbf{L},\label{eq:app:Y0knumeric}
\end{align}
where here the indices $l_k$ and $r_k$ indicate block-shifts associated to mode $k$.

The numerical implementation for the coherent-state ladder basis can be straightforwardly obtained by removing the parity indices $\mu$, $\nu$.

\section{Parametrizing the basis using higher-order processes}
\label{app:higherordersqueezing}
In addition to the parametrization of the basis by coherent-state amplitude $\alpha$ alone, we can include higher-order processes, such as squeezing and other higher-order processes. For a single bosonic mode, we define the following un-normalized state including up to $K$-photon moments,
\begin{equation}
       \ketb{\alpha^{(1)},\dots, \alpha^{(K)}} \equiv \prod\limits_{n=1}^K e^{\alpha^{(n)} {{}\hat{a}^\dagger}^n}  \ket{0}.
\end{equation}
We can naturally relate derivatives of the parameters $\alpha^{(n)}$ to the application of ${{}\hat{a}^\dagger}^n$:
\begin{equation}
    \pdv{\alpha^{(n)}} \ketb{\alpha^{(1)},\dots, \alpha^{(K)}} = 
    {{}\hat{a}^\dagger}^{n} \ketb{\alpha^{(1)},\dots, \alpha^{(K)}}.
\end{equation}
For a single bosonic mode, the equation of motion for $\dot{\alpha}^{(n)}$ then becomes
\begin{equation}
        \dot{\alpha}^{(n)} = [\boldsymbol{C^{-1}}\boldsymbol{Y}]_n,
\end{equation}
with
\begin{align}
    [\boldsymbol{C}]_{kl} &= \text{Tr}\{\boldsymbol{C_0}^{(k, l)}[\boldsymbol{B}\boldsymbol{S}\boldsymbol{B}]\},\\
    \boldsymbol{Y}_k &=  \text{Tr}\{\boldsymbol{Y_0}^{(k)} \boldsymbol{B}\},\\
    [\boldsymbol{C_0}^{(k,l)}]_{mn} &= \mel{\pdv{\phi_m}{\alpha^{(k)}}}{
    [\hat{\mathds{1}} - \hat{P} ]}{\pdv{\phi_n}{\alpha^{(l)}}}\\
    [\boldsymbol{Y_0}^{(k)}]_{mn} &= \mel{\pdv{\phi_m}{\alpha^{(k)}}}{[\hat{\mathds{1}} - \hat{P}] \mathcal{{L}}[\hat{\rho}]}{\phi_n}.
\end{align}

\end{document}